\documentclass[aps,prl,showpacs,twocolumn]{revtex4}
\usepackage{graphicx}
\usepackage[T1]{fontenc} 
\usepackage{epsfig,latexsym}
\usepackage{color}
\usepackage[latin1]{inputenc}
\usepackage{amsmath}
\usepackage{amssymb}
\usepackage{amsfonts}
\usepackage{indentfirst}
\usepackage{ae}
\usepackage{float}

\begin{document}
\title{Gouy Phase for Relativistic Quantum Particles}
\author{R. Ducharme$^{1}$ and I. G. da Paz$^{2}$}

\affiliation{$^{1}$ 2112 Oakmeadow Pl., Bedford, TX 76021, USA}

\affiliation{$^2$ Departamento de F\'{\i}sica, Universidade Federal
do Piau\'{\i}, Campus Ministro Petr\^{o}nio Portela, CEP 64049-550,
Teresina, PI, Brazil}

\begin{abstract}
Recently Gouy rotation was observed with focused non-relativistic
electron vortex beams. If the electrons in vortex beams are very
fast we have to take into account relativistic effects to
completely describe the Gouy phase on them. Exact Hermite-Gaussian 
solutions to the Klein-Gordon equation for particle beams are 
obtained here that make explicit the 4-position of the focal point 
of the beam. These are Bateman-Hillion solutions with modified phase 
factors to take into account the rest mass of the particles. They 
enable a relativistic expression for the Gouy phase to be determined. 
It is in fact shown all the solutions are form invariant under Lorentz 
transformations. It is further shown for the exact solutions to 
correspond to those of the Schr\"{o}dinger equation the relative time
between the focal point and any point in the beam must be constrained
to be a specific function of the relative spatial coordinates.
\end{abstract}


\maketitle

\textit{Introduction} - The Gouy phase of matter waves in the
non-relativistic context have been explored firstly in
\cite{PNP,Paz2}, followed by experimental realizations with
Bose-Einstein condensates \cite{cond}, electron vortex beams
\cite{elec2} and astigmatic electron matter waves using in-line
holography \cite{elec1}. Also, it was showed that to improve the
accuracy in imaging the dynamics of free-electron Landau states the
diffractive Gouy phase rotation has to be reduced \cite{landau}.

Electron vortex beams recently have been theoretically and
experimentally explored since they can be applied to improved
electron microscopy of magnetic and biological specimens
\cite{McMorran, Schattschneider}. Production of electron vortex
beams with high-energy and orbital quantum number seems feasible
\cite{Ivanov}. The Angular Momentum and Spin-Orbit interaction in
the relativistic electron vortex beams was studied in terms of the
Dirac equation in \cite{Nori}. The Gouy phase for this case, which
was not treated in \cite{Nori}, is completely explained only if
relativistic effects be considered and in this letter we explore
these possible effects in the Gouy phase. The experimentally
observed solutions for electron vortex beams with carry out Orbital
Angular Momentum (OAM) could be expressed in terms of the solutions
that will be obtained here. In order to avoid additional
complications we do not take into account the spin of particle and
we will solve here the Klein-Gordon equation.

Bateman \cite{HB} discovered a class of exact solutions to the
linear wave equation over a hundred years ago. Hillion \cite{PH,
KPC} later complexified these solutions for application to wave
packet and wave beam problems. We present exact Bateman-Hillion
solutions to the Klein-Gordon equation for the wave function
$\Psi(x_\mu)$ in Hermite-Gaussian beams where $x_\mu$ denotes
position in Minkowski space and $\mu = 0,1,2,3$. Here we
characterize the quantum dynamics of a relativistic particle
enabling us to better understand the quantum behavior of electrons in
high velocities, for example in relativistic electron vortex. As a
consequence of the relativistic effects the Gouy phase for the matter 
waves has a different expression from the non-relativistic counterpart
that must be taken into account, especially, when we focus on a 
relativistic particle beam. Also, we present a method for using these 
solutions to calculate the properties of beams for continuous wave 
sources.

The origin of the coordinate system for a beam is usually at the 
center of the focal point. It follows the spatial coordinates 
$x_i^r (i=1,2.3)$ of this point are zeroed out but translating the 
beam will make them explicit. For relativistic solutions, it must 
be recognized $x_i^r$ is part of a 4-vector such that the 4-potential 
is also dependent on the time coordinate $x_0^r$.

The fact $\Psi$ depends on two 4-position vectors creates a problem 
familiar from the treatment of two interacting relativistic particles 
\cite{AK, CA} that the field cannot evolve in two independent time 
coordinates. The known solution to be applied here is to use a Dirac 
delta function $\delta[f(\xi_\mu)]$ to impose a relationship $f(\xi_\mu)=0$ 
between the relative coordinates $\xi_\mu=x_\mu-x_\mu^r$. In classical 
electrodynamics, this idea leads to the derivation of the 
Lienard-Wiechert potentials \cite{SR} for the field experienced at one 
point owing to the presence of a point charge at another. The delta 
function notation for eliminating the relative time is also used in 
quantum electrodynamics \cite{RPF}.

For non-relativistic beams the wave function $\Psi$ can also be
calculated from the Schr\"{o}dinger equation \cite{PNP}. Solutions
to the Schr\"{o}dinger equation are harmonic in time but $|\Psi|$ is
independent of time. It is usual therefore to normalize $\Psi$ in a
3-dimensional constraint space rather than over all Minkowski space.
For the exact solutions to be developed in this letter $|\Psi|$ does
depend on time. The intent though is still to normalize them in a
3-dimensional constraint space using the $\delta[f(\xi_\mu)]$ condition 
to define it.

\textit{Hermite-Gaussian Beams} - Consider a beam of particles each
having a rest mass $m_0$, a 4-position $x_\mu = (x_i, t)$ and a
4-momentum $p_\mu = (p_i, E)$. The Klein-Gordon equation for the
wave function $\Psi (x_\mu)$ representing each of the particles in
Minkowski space can be expressed as
\begin{equation} \label{eq: KG}
(\hat{E}^2 + c^2\hat{p}_i^2)\Psi = m_0c^2\Psi
\end{equation}
where
\begin{equation} \label{eq: operators}
\hat{p_i} = \frac{\imath}{\hbar} \frac{\partial}{\partial x_i},
\quad \hat{E} = -\frac{\imath}{\hbar} \frac{\partial}{\partial t}
\end{equation}
are the momentum and energy operators, $\hbar$ is Planck's constant
divided by $2\pi$ and $c$ is the velocity of light.

Here we will find an exact solutions $\Psi_{mn}$ to the Klein-Gordon
equation (\ref{eq: KG}) for Hermite-Gaussian beams moving in the
$x_3$-direction. It will be assumed this takes the Bateman inspired
form
\begin{equation} \label{eq: bateman_solution}
\Psi_{mn} = \Phi_{mn}(\xi_1, \xi_2, \xi_3+c\tau) \exp[\imath(k_3 x_3 -\omega
t)],
\end{equation}
where 
\begin{equation} \label{eq: relative_coordinates}
\xi_i = x_i - x_i^r, \quad \tau = t -  t^r
\end{equation}
gives the position of each point $x_\mu$ relative to the center of the focal
point of the beam $x_\mu^r$, $k_3$ is the wave vector component along the 
direction of the beam, $\omega$ is the angular frequency and $\Phi_{mn}$ 
are scalar functions. The positive integers $m$ and $n$ indicate the mode of
the beam.

Inserting eq. (\ref{eq: bateman_solution}) into the Klein-Gordon
equation (\ref{eq: KG}) gives
\begin{equation} \label{eq: KG_bateman}
\left[ \frac{\partial^2}{\partial x_1^2} +
\frac{\partial^2}{\partial x_2^2} + 2 \imath \left( k_3
\frac{\partial}{\partial x_3} +
\frac{\omega}{c^2}\frac{\partial}{\partial t} \right) \right]
\Phi_{mn} = 0,
\end{equation}
having spotted
\begin{equation}
\left( \frac{\partial^2}{\partial x_3^2} - \frac{1}{c^2}
\frac{\partial^2}{\partial t^2}\right) \Phi_{mn} = 0,
\end{equation}
owing to the dependence of $\Phi_{mn}$ on $\xi_3$ and $\tau$ in the
linear combination $\xi_3+c\tau$.

Equation (\ref{eq: KG_bateman}) is analogous to the paraxial wave
equation
\begin{equation} \label{eq: paraxial_wave_equation}
\frac{\partial^2\Phi_{mn}^P}{\partial x_1^2} +
\frac{\partial^2\Phi_{mn}^P}{\partial x_2^2} + 2 \imath k_3
\frac{\partial \Phi_{mn}^P}{\partial x_3}   = 0,
\end{equation}
where the role of $\xi_3$ in $\Phi^P(\xi_1, \xi_2, \xi_3)$ is replaced by
$\xi_3+c\tau$ in $\Phi(\xi_1, \xi_2, \xi_3+c\tau)$.  It can therefore be 
solved analogously \cite{AES} to give
\begin{eqnarray} \label{eq: hermite_gauss_solution}
\Phi_{mn} = \frac{C_{mn} w_0}{w}H_m\left(
\frac{\sqrt{2}\xi_1}{w}\right) H_n\left(
\frac{\sqrt{2}\xi_2}{w}\right)\times\\ \exp \left[ \frac{\imath
(k_3+\frac{\omega}{c}) (\xi_1^2+\xi_2^2)}{\xi_3+c\tau-2\imath b} - \imath
g_{mn} \right],
\end{eqnarray}
where $H_m$ and $H_n$ are Hermite polynomials,
\begin{equation} \label{eq: b_term}
b = \frac{1}{2}\left( k_3 + \frac{\omega}{c}\right) w_0^2,
\end{equation}
\begin{equation} \label{eq: spot_radius}
w(\xi_3,\tau) = w_0\sqrt{1+\left( \frac{\xi_3+c\tau}{2b} \right)^2},
\end{equation}
is the beam radius, $w_0=w(0,0)$ is the beam waist and
\begin{equation} \label{eq: gouy_phase}
g_{mn}(\xi_3,\tau) = (1+m+n)\arctan \left( \frac{\xi_3+c\tau}{2b} \right),
\end{equation}
is the Gouy phase of a relativistic quantum particle. We observe
that it depends of the longitudinal direction $x_3$ as the
non-relativistic case but also it depends of the time and the speed
of light. The Gouy phase is a necessary feature of our solutions
that are relativistic and can be normalized. We observe that the
relativistic effect in the Gouy phase can be negligible only if the
evolution occur in a scale less than $1\;\mathrm{ns}$ and the
longitudinal distance is large than $1\;\mathrm{m}$.

It is proposed the normalizing constant $C_{mn}$ in $\Phi_{mn}$ can
be determined from the condition
\begin{equation} \label{eq: normalization}
\int_{-\infty}^{+\infty} |\Psi_{mn}|^2 \delta[f(\xi_\mu)]d^4\xi = 1,
\end{equation}
where $d^4\xi = d\xi_1d\xi_2d\xi_3d\tau$ and $f(\xi_\mu)$ is to be 
determined. For physical interpretation, it is clear a physical system 
cannot evolve in the two independent times $t$ and $\tau$. The intent
of the $f(\xi_\mu)=0$ constraint condition is therefore to eliminate 
the relative time in terms of the relative spatial coordinates.


\textit{Lorentz Transformations} - The wave function
$\Psi_{mn}(x_\mu)$ has been determined in eq. (\ref{eq:
hermite_gauss_solution}) to be an exact solution of the Klein-Gordon
eq. (\ref{eq: KG}) for a particle in a beam. The task ahead is to
confirm these solutions are form preserving under Lorentz
transformations.

A 4-vector $q_\mu$ is relativistic if it preserves its form under
the Lorentz transformation equations:
\begin{equation} \label{eq: lorentz_trans_x}
q_i^{\prime} = q_i + \gamma \frac{v_i}{c} \left(\frac{\gamma
}{1+\gamma}\frac{v_jq_j}{c} - q_0 \right),
\end{equation}
\begin{equation} \label{eq: lorentz_trans_t}
q_0^{\prime} = \gamma(q_0-\frac{v_j q_j}{c}),
\end{equation}
(see ref. \cite{SR}) where $v_i$ is the relative velocity between
any two inertial reference frames and $\gamma = (1-v^2)^{-1/2}$. For
current purposes $q_\mu$ belongs to the set of position $x_\mu$, 
relative position $\xi_\mu$, wave vector $k_\mu$ and 4-momentum 
$p_{\mu}$.

To confirm the Hermite-Gaussian beam solutions are fully
relativistic it will be necessary to investigate the Lorentz
transformation of number of different quantities including the phase
factor, the Gaussian function, the Hermite functions, and Gouy
phase. For brevity, the analysis will be limited to the case of an
observer that is moving parallel to the axis of the beam.

The product $k_\mu x_\mu$ is Lorentz covariant implying
\begin{equation} \label{eq: phase_factor}
k_\mu^\prime x_\mu^\prime = k_\mu x_\mu.
\end{equation}
The numerator and denominator in the Gaussian component of eq.
(\ref{eq: hermite_gauss_solution}) can be transformed separately to
give
\begin{equation} \label{eq: gaussian_numerator}
\left(k_3^\prime+\frac{\omega^\prime}{c} \right) (\xi_1^{\prime 2} +
\xi_2^{\prime 2}) = \sqrt{\frac{c-v}{c+v}} \left(k_3+\frac{\omega}{c}
\right) (\xi_1^{2} + \xi_2^{2}),
\end{equation}
and
\begin{equation} \label{eq: gaussian_denominator}
\xi_3^{\prime} + c\tau^{\prime} - 2\imath b^{\prime} =
\sqrt{\frac{c-v}{c+v}} (\xi_3 + c\tau - 2\imath b),
\end{equation}
having determined
\begin{equation} \label{rayliegh_transform}
b^{\prime} = \sqrt{\frac{c-v}{c+v}} b,
\end{equation}
from eq. (\ref{eq: b_term}) assuming the radius $w_0$ is a Lorentz
invariant scalar. Putting these results together leads to
\begin{equation} \label{eq: gaussian_argument}
\frac{\left(k_3^\prime+\frac{\omega^\prime}{c} \right) (\xi_1^{\prime
2} + \xi_2^{\prime 2})}{\xi_3^{\prime} + c\tau^{\prime} - 2\imath
b^{\prime}} = \frac{\left(k_3+\frac{\omega}{c} \right) (\xi_1^{2} +
\xi_2^{2})} {\xi_3 + c\tau - 2\imath b}.
\end{equation}
Eqs. (\ref{eq: gaussian_numerator}) through (\ref{eq:
gaussian_argument}) can now be used to transform eq. (\ref{eq:
hermite_gauss_solution}) into $\Phi_{mn}^\prime$ is equal to
\begin{eqnarray} \label{eq: hermite_gauss_solution_moving}
\frac{C_{mn} w_0}{w^{\prime}}H_m\left(
\frac{\sqrt{2}\xi_1^{\prime}}{w^{\prime}}\right) H_n\left(
\frac{\sqrt{2}\xi_2^{\prime}}{w^{\prime}}\right)\times\\
\exp \left[ \frac{\imath \left(k_3^\prime+\frac{\omega^\prime}{c}
\right) (\xi_1^{\prime 2}+\xi_2^{\prime
2})}{\xi_3^{\prime}+c\tau^{\prime}-2\imath b^{\prime}} - \imath
g_{mn}^{\prime} \right],
\end{eqnarray}
where
\begin{equation}
w^{\prime} = w_0\sqrt{1+\left(
\frac{\xi_3^{\prime}+c\tau^{\prime}}{2b^{\prime}} \right)^2} =
w_0\sqrt{1+\left( \frac{\xi_3+c\tau}{2b} \right)^2},
\end{equation}
\begin{equation}
\frac{g_{mn}^{\prime}}{(1+m+n)} = \arctan \left(
\frac{\xi_3^{\prime}+c\tau^{\prime}}{2b^{\prime}} \right) = \arctan
\left( \frac{\xi_3+c\tau}{2b} \right),
\end{equation}
and $C_{mn}$ is a Lorentz invariant scalar.

It follows from taking stock of all of these results that eq.
(\ref{eq: bateman_solution}) can be rewritten as
\begin{equation} \label{eq: solution_hermite_moving}
\Psi_{mn}^{\prime} = \Phi_{mn}(\xi_1^{\prime}, \xi_2^{\prime},
\xi_3^{\prime}+c\tau^{\prime}) \exp (\imath k_3^{\prime} x_3^{\prime}
-\imath \omega^{\prime} t^{\prime}).
\end{equation}
On comparing eqs. (\ref{eq: bateman_solution}) and (\ref{eq:
solution_hermite_moving}) it is concluded $\Psi_{mn}$ is form
invariant under Lorentz transformations and therefore fully
relativistic.


\textit{Correspondence to the Schr\"{o}dinger Equation} - The
objective here is to determine the form of the constraint condition
$f(\xi_\mu) = 0$ in eq. (\ref{eq: normalization}) from the principle
that exact solutions of the Klein-Gordon equation should correspond
to solutions of the Schr\"{o}dinger equation in the non-relativistic
limit. The Schr\"{o}dinger equation takes the form
\begin{equation} \label{eq: schrodinger_wave_equation}
\frac{\partial^2 \Psi_{mn}^S}{\partial x_1^2} + \frac{\partial^2
\Psi_{mn}^S}{\partial x_2^2} + \frac{\partial^2
\Psi_{mn}^S}{\partial x_3^2} + 2 \imath \frac{m}{\hbar}
\frac{\partial \Psi_{mn}^S}{\partial t} = 0.
\end{equation}
For paraxial beams, the solution $\Psi_{mn}^S$  may be assumed to
take the form
\begin{equation} \label{eq: schrodinger_solution}
\Psi_{mn}^S = \Phi_{mn}^S(\xi_1, \xi_2, \tau) \exp
\left[\frac{\imath}{\hbar}(p_3 x_3 -E_s t) \right],
\end{equation}
where
\begin{equation} \
E_s = \frac{p_3^2}{2m_0},
\end{equation}
is non-relativistic kinetic energy. Inserting eq. (\ref{eq:
schrodinger_solution}) into eq. (\ref{eq:
schrodinger_wave_equation}) gives
\begin{equation} \label{eq: schrodinger_phi_equation}
\frac{\partial^2 \Phi_{mn}^S}{\partial x_1^2} + \frac{\partial^2
\Phi_{mn}^S}{\partial x_2^2} + 2 \imath \frac{m}{\hbar}
\frac{\partial \Phi_{mn}^S}{\partial t} = 0.
\end{equation}
Equation (\ref{eq: schrodinger_phi_equation}) is analogous to eq.
(\ref{eq: KG_bateman}) where the role of $\tau$ in $\Phi^S(\xi_1, \xi_2,
\tau)$ replaces $\xi_3+c\tau$ in $\Phi(\xi_1, \xi_2, \xi_3+c\tau)$.

The solution to eq. (\ref{eq: schrodinger_phi_equation}) is
\begin{eqnarray} \label{eq: schrodinger_phi_solution}
\Phi_{mn}^S = \frac{C_{mn} w_0}{w^S}H_m\left(
\frac{\sqrt{2}\xi_1}{w^S}\right) H_n\left(
\frac{\sqrt{2}\xi_2}{w^S}\right)\times\\
\exp \left[- \frac{m_0 (\xi_1^2+\xi_2^2)}{m_0w_0^2+2\imath \hbar \tau} -
\imath g_{mn}^S \right],
\end{eqnarray}
where
\begin{equation} \label{eq: schrodinger_spot_radius}
w^S(\tau) = w_0\sqrt{1+\left( \frac{\hbar \tau}{m_0w_0^2} \right)^2},
\end{equation}
is the beam radius and
\begin{equation} \label{eq: schrodinger_gouy_phase}
g_{mn}(\tau) = (1+m+n)\arctan \left( \frac{\hbar \tau}{m_0w_0^2} \right),
\end{equation}
is the non-relativistic Gouy phase for matter waves.

It can now be shown that the Schr\"{o}dinger and exact Klein-Gordon
beam solutions are related to each other through the expression
\begin{equation} \label{eq: bateman_schrodinger_connection}
\Phi_{mn}^S = \int \Phi_{mn} \delta(\xi_3-u_3\tau) dt,
\end{equation}
in the non-relativistic limit. This result is most easily
demonstrated for the case of a Gaussian beam
\begin{equation} \label{eq: gaussian_solution_KG}
\Phi_{00} = \frac{C_{00}b}{b+\imath \frac{1}{2}(\xi_3+c\tau)} \exp \left[
\frac{\imath (k_3+\frac{\omega}{c}) (\xi_1^2+\xi_2^2)}{\xi_3+c\tau-\imath 2b}
\right].
\end{equation}
On setting,
\begin{equation}
k_3 = \frac{m_0u_3 \gamma}{\hbar}, \quad \omega = \frac{m_0c^2
\gamma}{\hbar},
\end{equation}
eq. (\ref{eq: bateman_schrodinger_connection}) evaluates to
\begin{equation} \label{eq: gaussian_solution_schrodinger}
\Phi^S_{00} = \frac{C_{00}m_0w_0^2}{m_0w_0^2+2\imath \hbar \tau} \exp
\left[ \frac{-m_0 (\xi_1^2+\xi_2^2)}{m_0w_0^2+2\imath \hbar \tau}  \right],
\end{equation}
having assumed $\gamma \simeq 1$. The calculation for all the higher
order modes follows similarly.

It has been shown that in order for exact Bateman-Hillion solutions
of the Klein-Gordon equation for Hermite-Gaussian beams of matter
waves to correspond to solutions of the Schr\"{o}dinger equation in
the non-relativistic limit, they must be restricted to the
constraint space
\begin{equation} \label{eq: paraxial_constraint}
f(\xi_\mu) = \xi_3 - u_3\tau = 0.
\end{equation}
This result is straightforward to interpret for particles large
enough to have classical properties since the expression $\xi_3-u_3\tau =
0$ describes the trajectory of a classical free particle along the
axis of the beam. It is unexpected that smaller quantum particles
also appear to be confined to this constraint space.


\textit{Discussion} - Exact solutions have been derived to the
Klein-Gordon equation for matter waves Hermite-Gaussian beams using
Bateman-Hillion functions. The solutions have been shown to preserve
their form under Lorentz transformations. These exact solutions are
not time-harmonic but it has been found they can be used to produce
results comparable to time-harmonic solutions providing they are
normalized and interpreted in a 3-dimensional constraint space. The
guiding principle is time-harmonic and Bateman-Hillion solutions can
only be compared if the calculations are performed in spaces that
have the same number of dimensions.

It has been shown that for Klein-Gordon and Schr\"{o}dinger beams to
be in correspondence in the non-relativistic limit, the Klein-Gordon
solution must be restricted to the $\delta(\xi_3-u_3\tau)$ constraint
space. The constraint equation $\xi_3-u_3\tau = 0$ describes the
trajectory of a classical particle. In this sense, it is intuitive
that particles large enough to exhibit classical properties would
belong to this constraint space but surprising that smaller quantum
particles also appear to be restricted to it. When the Gouy phase of
matter waves was obtained in \cite{PNP}, the constraint equation
$\xi_3-u_3\tau = 0$ was used to transform the Schr\"{o}dinger equation
analogous to paraxial equation for which the propagation direction
of a quantum particle is said to be classical. As the energy
associated with the momentum of the particles in the propagation
direction is very high to include the relativistic effects we can
consider, as in \cite{PNP}, a classical movement in this direction
which makes the Gouy phase be a function of propagation direction and
the relativistic effect appear in the time coordinate, see eq.
\eqref{eq: gouy_phase}. If we do not consider that the momentum of
the particles in the propagation direction is high we have the
non-relativistic Schr\"{o}dinger equation that is different from
the paraxial equation but still the Gouy phase is present in the
solutions of a free particle as a property of the wave behavior.
However, it is a time dependent function instead of a function
dependent of the propagation direction.

In conclusion, we have shown that the Gouy phase of relativistic
quantum particle can be obtained by transforming the Klein-Gordon
equation to an analogue of paraxial equation which provides normalized
and Lorentz invariant solutions. The relativistic Hermite-Gaussian 
solutions contain Orbital Angular Momentum (OAM) 
and form a complete basis. These results can therefore be applied to 
treat relativistic electron vortex beams. It is anticipated though for
strongly converging or diverging beams that $\delta(\xi_3-u_3\tau)$
would need to be replaced to a form such as $\delta(|\xi_i|-|u_i|\tau)$
for better correspondence to the trajectories of classical particles 
in the case $m_0$ is large.

\end{document}